\documentclass[12pt,elsart,epsfig]{article}     
\usepackage{epsfig}
\begin{document}

\begin{center}
{\Large \bf Observation of $f_0(1770) \to \eta \eta$ in $\bar pp \to \eta\eta\pi^0$
reactions from 600 to 1200 MeV/c}
\vskip 5mm
{A.V. Anisovich$^d$, C.A. Baker$^a$, C.J. Batty$^a$, D.V. Bugg$^b$, R.P. Haddock$^c$,
C. Hodd$^b$, V.A. Nikonov$^d$, A.V. Sarantsev$^d$, V.V. Sarantsev$^d$, I. Scott$^b$, 
B.S.~Zou$^{b}$ \footnote{Now at IHEP, Beijing 100039, China} \\
{\normalsize $^a$ \it Rutherford Appleton Laboratory, Chilton, Didcot OX11 0QX,UK}\\
{\normalsize $^b$ \it Queen Mary and Westfield College, London E1\,4NS, UK}\\
{\normalsize $^c$ \it University of California, Los Angeles, CA90024, USA}\\
{\normalsize $^d$ \it PNPI, Gatchina, St. Petersburg district, 188350, Russia}\\ 
[3mm]}
\end {center}

\begin{abstract}
We present data on $\bar pp \to \eta \eta \pi ^0$ at beam momenta of 600, 900,
1050, and 1200  MeV/c.
At the higher three momenta, a signal is clearly visible due to 
$\bar pp \to f_0(1770)\pi ^0$, $f_0(1770) \to \eta \eta$.
It has mass $1770 \pm 12$ MeV and width $220 \pm 40$ MeV, where errors cover
systematic uncertainties as well as statistics. 

\end{abstract}

\vskip 5mm

We investigate $\bar pp \to \eta \eta \pi ^0$ at $\bar p$ momenta from 600 to
1200 MeV/c.
Data were taken at LEAR by the Crystal Barrel Collaboration.
Data from 1350 to 1940 MeV/c are reported in an accompanying paper \cite {1}.
They provide evidence for $f_0(2100) \to \eta \eta$, $a_2(1660) \to \eta \pi$
and a broad $f_2(1980) \to \eta \eta$.
The first resonance lies above the mass range of present data, but is an
important element in $0^+$ spectroscopy.
The $a_2(1660)$ is strong at higher momenta but, because its threshold is at
$\sim 2200$ MeV, it makes only a minor contribution to present data up to 
1200 MeV/c.
The $f_2(1980)$ plays an important role, since it is broad and extends into the
present mass range.
We shall provide evidence for an $I=0$, $J^{PC} 0^{++}$ resonance at 1770 MeV,
decaying to $\eta \eta$. 
This resonance is important in trying to unravel the mysteries concerning the
$f_J(1710)$ \cite {2}, hence in establishing the systematics of $0^+$ mesons.

At lower masses, there has been earlier evidence from the Crystal Barrel
collaboration for an $I=0$ resonance whose mass is variously estimated from
1300 to 1380 [3-6].
There is also an $I=1$ resonance at 1450 MeV \cite {7} and finally the well
known $f_0(1500)$.
At higher mass, the WA102 collaboration has published evidence for an
$f_0(2020)$ resonance decaying to $4\pi$ \cite {8}.

\begin {table}[htp]
\begin {center}
\begin {tabular}{ccc}
\hline
Momentum & Events & Background\\
(MeV/c)  &        & ($\%$) \\\hline
600 & 2922 & 2.8 \\
900 & 9023 & 3.9\\
1050& 6607 & 3.3\\
1200& 9594 & 3.7\\\hline
\end {tabular}
\caption {Numbers of events and estimated background levels}
\end {center}
\end {table}

The experimental conditions and data processing are identical to those of the
higher momenta reported in Ref. \cite {1}, and we refer to that paper for
experimental details.
The largest backgrounds are from $\eta \pi ^0\pi ^0\pi ^0$ and $\omega \pi^0 \pi^0$,
two photons being lost in the former case and one in the latter.
Numbers of events and estimated backgrounds are shown in Table 1.
Using cross sections observed in these other channels, we find that backgrounds
are close to a phase space distribution. 
They are included in fitting the data, but omitting them has negligible effect
on the analysis, where log likelihood changes by $<4$.

\begin{figure}
\begin{center}
\epsfig{file=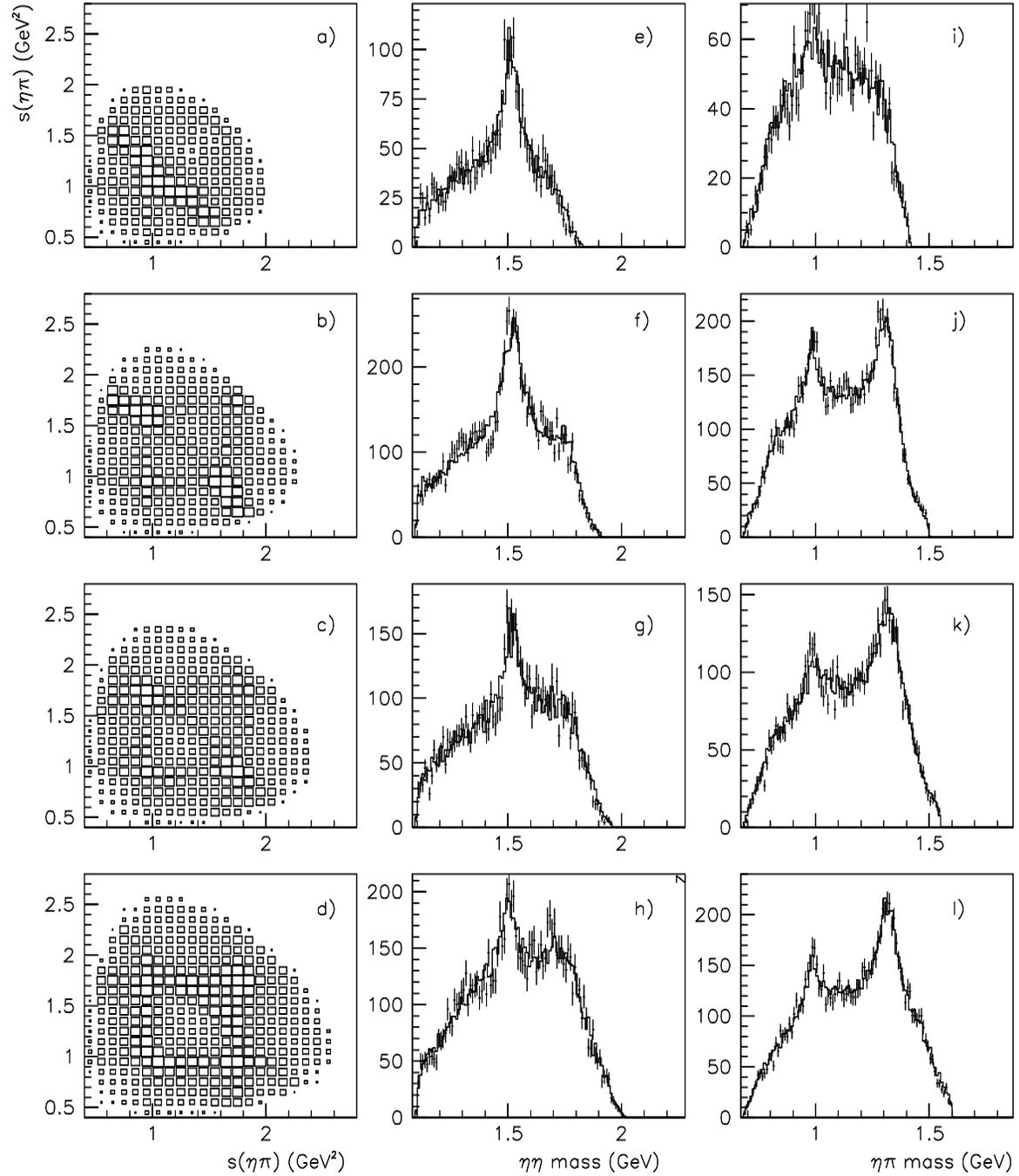,width=17cm}\
\vskip -19.897cm
\epsfig{file=17NEW.PS,width=17cm}\
\vskip -8mm
\caption {(a)-(d) Dalitz plots at 600, 900, 1050 and 1200 MeV/c,
(e)-(h) Projections on to $M(\eta \eta)$,
(i)-(l) Projections on to $M(\eta \pi )$.}
\end{center}
\end{figure}

Fig. 1 shows Dalitz plots at the four beam momenta and mass projections on to
$\eta \eta$ and $\eta \pi$.
Histograms superimposed on the projections are the result of the maximum
likelohood fit described below.
There are conspicuous peaks due to $f_0(1500)$, $a_0(980)$ and $a_2(1320)$.
The feature of interest is a shoulder at $\sim 1770$ MeV in $\eta \eta$,
clearly visible in Figs. 1(f), (g) and (h) at 900, 1050 and 1200 MeV/c.
It is barely visible at the higher momenta reported in Ref. [1].
At 1350 MeV/c, there is a shoulder at $\sim 1770$ MeV, and the amplitude
analysis gives a small optimum there, but the significance is found to be
much less than at lower momenta.

We shall report here a full partial wave analysis of production and decays
of resonances.
Amplitudes we find to be significant are listed in Table 2.
In making this choice, we are guided by experience that it requires roughly
225 MeV/c of momentum per unit of orbital angular momentum $L$ in the
production process.
Present data are consistent with this guideline.
The last column of Table 2 shows momenta in the decay process at the highest
beam momentum of 1200 MeV/c.
For the $a_0(980)\eta$ channel, we require up to $L=3$.
There is no evidence for the presence of $L=4$;
this agrees with the expectation that the first $4^-$ $q\bar q$ state will
lie at a mass of 2300 MeV or above, i.e. a beam momentum of at least 1650
MeV/c.
The $a_0(980)$ is fitted with a Flatt\' e form using parameters derived from
earlier data on $\bar pp \to \eta \pi ^0 \pi ^0$ at rest \cite {9}.
The $a_0(980)$ signal is slightly stronger in the data than the fit,
particularly in Fig.1(i); the fit can be improved by reducing the width of the
$a_0(980)$ by $\sim 20\%$.
The $f_0(1500)\pi$ channel is described successfully with waves up to $L=2$.
The strong $a_2(1320)\eta$ channel requires up to $L=3$, but again there is no
evidence for $L=4$; the $2^+$ initial state with helicity 0 makes quite a
significant $L=3$ contribution: $(4-8)\%$ at 900 MeV/c and above.
 
\begin {table}[htp]
\begin {center}
\begin {tabular}{ccc}
\hline
Channel  & Partial waves  & CM momentum\\
         &                & (MeV/c)   \\\hline
$a_0(980)\eta$ & $^1S_0$,$^3P_1$,$^1D_2$,$^3F_3$ & 734 \\
$f_0(1500)\pi$ & $^1S_0$,$^3P_1$,$^1D_2$         & 539 \\
$a_2(1320)\eta$ & $^1D_2(L=0$ and 2),$^1S_0(L=2)$ & 498 \\
 &$^3P_1$, $2^+(m=0$ and 1), $^3F_3(L=1)$ &     \\
& $2^+(L=3,m=0$) &     \\
$f_0(1770)\pi$ & $^1S_0,3P_1$ & 323 \\
$f_2(1270)\pi$ & $^1D_2$      & 690 \\
$f_2(1980)\pi$ & $^1D_2,^3P_1,2^+(m=0$ and 1),$^3F_3$ & 100\\\hline
\end {tabular}
\caption {Partial waves included in the analysis; the third column
shows the centre of mass momentum on resonance for the highest beam
momentum of 1200 MeV/c}
\end {center}
\end {table}

Figs. 2 and 3 show examples of angular distributions in bands centred on resonances.
The fits, shown by histograms, are equally good at other momenta.
Fig. 2 shows the production angular distribution in terms of the centre of mass
angle $\tau$ at which the resonance is produced.
Fig. 3 shows decay angular distributions of a resonance, e.g. $a_2(1320) \to
\eta \pi$, in terms of its decay angle $\alpha$ with respect to the beam direction.
\begin{figure}
\begin{center}
\vskip -15mm
\epsfig{file=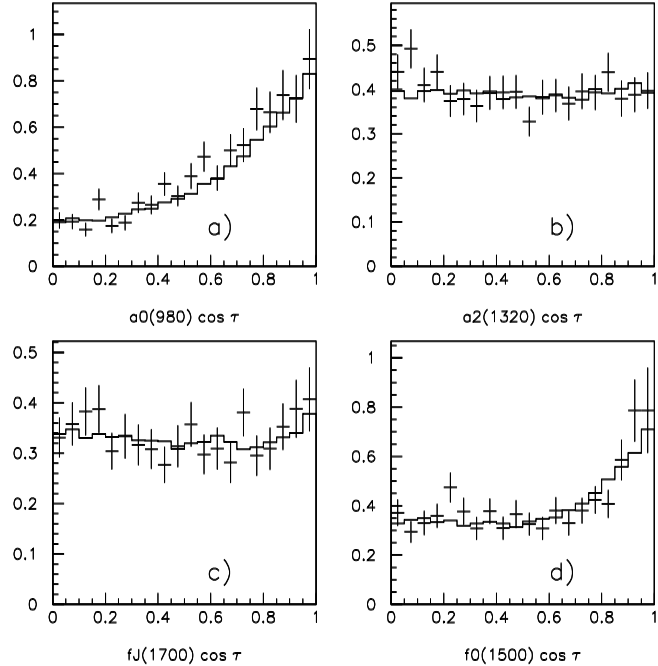,width=10cm}\
\vskip -10.047cm
\epsfig{file=B1050.PS,width=10cm}\
\vskip -4mm
\caption {A comparison of the fit at 1200 MeV/c with angular distributions
(unnormalised) against $\cos \tau$, where $\tau$ is the centre of mass angular
production angle; (a) a 65 MeV wide band around $a_0(980)$, (b) a 118 MeV wide
band around $a_2(1320)$, (c) an $\eta \eta$ band from 1640 to 1840 MeV, 
(d) a 120 MeV wide band around $f_0(1500)$. Points with errors are data and
the histogram shows the fit. }
\end{center}
\end{figure}

Our data on the $3\pi ^0$ state (not shown here) contain strong $f_2(1270)\pi ^0$
signals and lead us to expect weak $f_2(1270) \to \eta \eta$ contributions at the
level of $3-5\%$ in present data; these are barely visible in the projections.
We find that this weak channel can be approximated with $L=0$ production only;
this approximation has no impact on the interesting $\eta \eta $ mass range
around 1770 MeV.
We have tried including $f_0(1370)$ in the fit, but no significant contribution is
required.

We find that most partial waves follow within errors a smooth dependence on centre
of mass momentum, but with fluctuations which are typically $25\%$ for the larger
waves ($\sim 8\%$ branching ratio) and $50\%$ for smaller waves contributing
branching ratios of only $2\%$.
Contributions below $1\%$ are dropped.
We have data on $\bar pp \to \eta \pi ^0 \pi ^0$ with roughly a factor 12--16
higher statistics \cite {10}.
There, many resonances are visible, but phase variations with momentum are mostly
small, i.e. resonances are phase coherent.
In present data, phase variations are again small, although there are variations of
typically $\pm (10-25)^\circ$ from one momentum to another.
The fit is made by the maximum likelihood method. 
Our definition is such that a change of log likelihood of 0.5 corresponds to one
standard deviation for a change of one parameter.
The acceptance of the detector is simulated by Monte Carlo events, processed through
exactly the same selection procedure as data; statistics of accepted Monte Carlo
events are roughly 4 times those of data events.

\begin{figure}
\begin{center}
\vskip -10mm
\epsfig{file=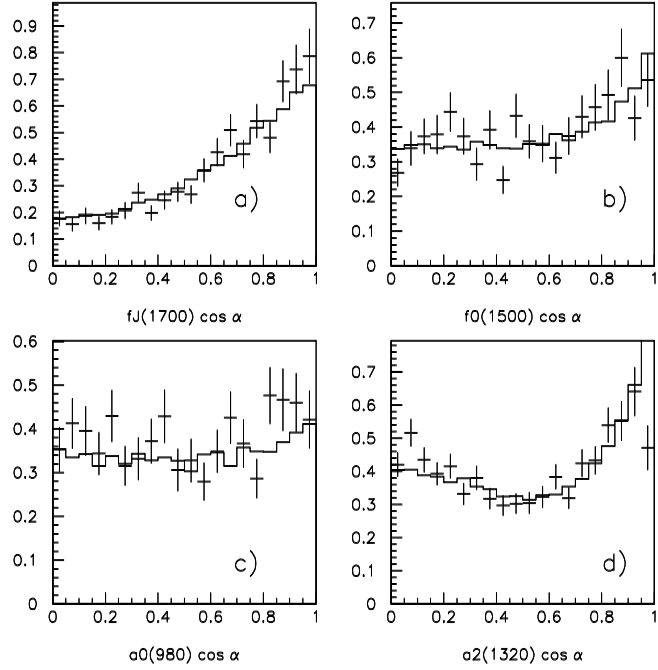,width=10cm}\
\vskip -10.047cm
\epsfig{file=C1050.PS,width=10cm}\
\vskip -4mm
\caption {As Fig. 2 for decay angular distibutions with respect to the beam in the
rest frame of the resonance. }
\end{center}
\end{figure}
The fitting algorithm is sufficiently fast that the amplitude analysis produces a
fit in roughly 60 seconds of computing.
This is sufficiently fast that we have been able to examine a large number
$(> 1500)$ of trials with varying ingredients in order to study the systematics of
the fit, its stability and sensitivity to every channel.
Although some ambiguities are present at individual momenta (particularly between
$L=0$ and $L=2$ and between $L=1$ and $L = 3$), we have located only one solution
having a smooth variation of magnitude, and particularly phase, with momentum.
The Dalitz plots of Fig. 1 are reproduced so closely by the fit that no difference
from data is visible within statistics.

\begin{figure}
\begin{center}
\vskip -20mm
\epsfig{file=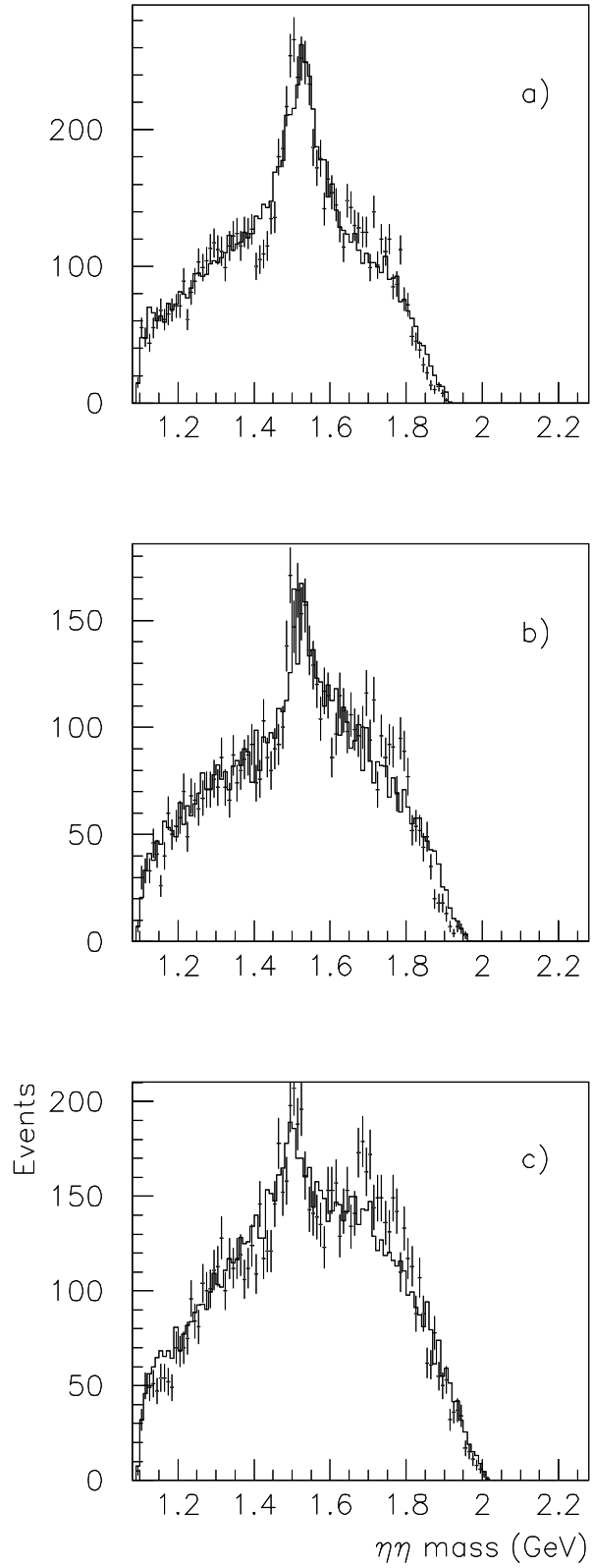,width=10cm}\
\vskip -23.832cm
\epsfig{file=NO1770.PS,width=10cm}\
\vskip -8mm
\caption {Fits without any $f_J(1770)$ at beam momenta of (a) 900, (b) 1050 and
(c) 1200 MeV/c. }
\end{center}
\end{figure}
The $\eta \eta$ signal at 1770 MeV is visible in $\eta \eta$ projections of
Figs. 1(f), (g) and (h) at 900, 1050 and 1200 MeV/c.
Fits without any resonance in this mass range are shown in Figs. 4(a)-(c).
There is an obvious surplus of events between 1700 and 1800 MeV.
We have tried fitting it with $J=0$ or 2.
Variations of log likelihood with its mass are shown in Figs.5(a)-(c).
There is a strong, well defined optimum for $J=0$ (full curve).
Optimum masses are $1758 \pm 10$ MeV at a beam momentum of 900 MeV/c,
$1770 \pm 8$ at 1050 MeV/c and $1775 \pm 8$ MeV at 1200 MeV/c.
Branching fractions are respectively $5.5\%$, $11.7\%$ and $10.1\%$ at these
three momenta.
At 1350 MeV/c, there is a shallow minimum at the same mass and width, but the
fitted $f_0(1770)$ contribution is only $1.5\%$.
Data at 600 MeV/c are too close to threshold to show an optimum in the mass or
width of $f_0(1770)$.
Column (a) of Table 3 lists changes in log likelihood when $f_0(1770)$ is omitted
from the fit.

\begin {table}[htp]
\begin {center}
\begin {tabular}{ccccc}
\hline
Momentum & (a) & (b) & (c) & (d) \\
(MeV/c)  &     &     &     &     \\\hline
900 & -81.4   & -33.8& 16.0  & 3.1\\
1050& -101.6  & -34.2& 19.3  & 4.8\\
1200& -83.0   & -28.9& 17.5  & 3.5\\
1350& -7.7    & -5.4 & 19.2  & 3.4\\\hline
\end {tabular}
\caption {Changes to log likelihood of the basic fit, (a) dropping $f_0(1770)$,
(b) replacing $f_0(1770)$ by $f_2(1770)$, (c) adding $f_2(1712)$ or (d) $f_0(1712)$}
\end {center}
\end {table}

Spin 0 gives a better optimum than $J=2$ (shown by dashed curves in Fig.5).
Differences of log likelihood are given in column (b) of Table 3.
However, spin 2 still shows some optimum at a similar mass to spin 0.
We find this is due to feed-through from mis-identified spin 0 signal.
The detector has holes around the beam with an opening angle of $12^\circ$.
This incomplete coverage allows some cross-talk between spin 0 and spin 2.
In order to examine this, we have adopted two simulation procedures.
In the first, we have used Monte Carlo events to generate data samples
with the same partial wave amplitudes as are fitted to $f_0(1770)$ plus other
channels. Then we have fitted these samples replacing $f_0(1770)$ by a 
$2^+$ resonance.
This simulation reproduces the magnitudes of the dashed curves of Fig. 5(a) and (b)
at 900 and 1050 MeV/c quite well.

\begin{figure}
\begin{center}
\vskip -12mm
\epsfig{file=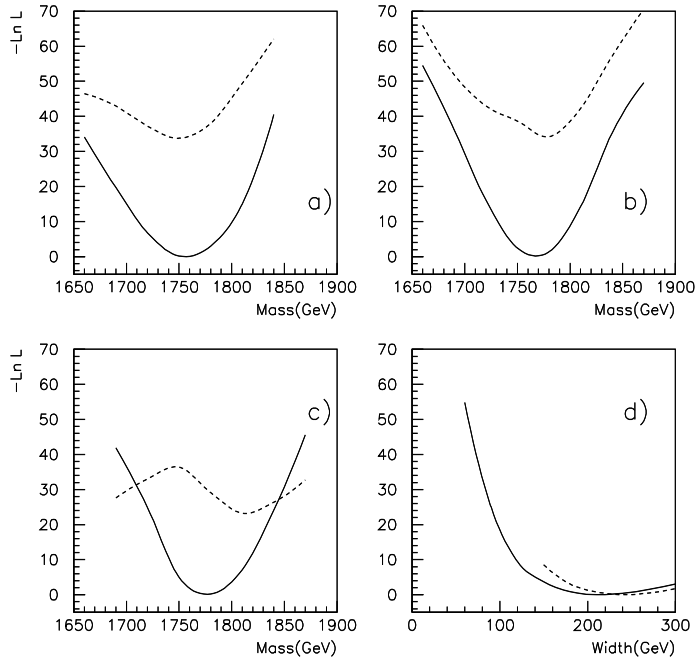,width=10cm}\
\vskip -10.037cm
\epsfig{file=4F0177.PS,width=10cm}\
\vskip -6mm
\caption {Variations of log likelihood with the mass of $f_0(1770)$, full line
$J=0$, dashed line $J=2$, (a) at 900 MeV/c, (b) at 1050 MeV/c, (c) at 1200 MeV/c'
(d) the variation of log likelihood with width at 1050 MeV/c (full curve) and
1200 MeV/c (dashed). }
\end{center}
\end{figure}

The fits with spin 2 use five partial waves $^1D_2$, $^3P_1$, $2^+(m=0$ and 1) and
$^3F_3$, while fits with spin 0 use two partial waves $^1S_0$ and $^3P_1$.
Statistically, we would expect the fits with spin 2 to be better than those with
spin 0 by 3 in log likelihood, because of the 6 extra fitting parameters.
In practice, we find this to be an under-estimate, since certain combinations
of amplitudes producing a $J=2$ resonance with $L=0$ and 1 can simulate closely
the flat angular distribution of the dominant $^1S_0$ amplitude producing $J=0$.
To examine this, our second simulation is to fit fake $J=2$ resonances of width
220 MeV to the simulated data sets.
What we observe in 10 such trials is that the fake $2^+$ signal improves log
likelihood by an average of $17 \pm 3$.
This sets a scale of what may be fitted incorrectly by spin 2; any improvement
smaller than this cannot be believed.
Our conclusion is that a fake $2^+$ signal produced with $L=0$ or 1 from 5 initial
partial waves can be expected to fit the data better than a fake $J=0$, produced
from 2 initial partial waves, by approximately 12 in log likelihood.
If we use this result, the conclusion is that $J=0$ is preferred over $J=2$ by
$\sim 46$ in log likelihood at 900 MeV/c, by $\sim 46$ at 1050 MeV/c and by
$\sim 41$ at 1200 MeV/c.
These are statistically each over 9 standard deviations.
If, pessimistically, we ignore the difference in the number of fitted partial waves
and use just raw differences in log likelihood between spin 0 and 2, the discrimination
is still over 8 standard deviations at each momentum.

Fig. 5(d) shows log likelihood versus the width of the $f_0(1770)$ at 1050 and
1200 MeV/c, where the data are most definitive.
At other beam momenta, the optimum is somewhat less well defined and flattens out
at large widths.
From our studies with a variety of ingredients, we estimate that the systematic errors
on the mass and width of the resonance are respectively $\pm 12$ and $\pm 40$ MeV.

There has been previous evidence for $f_0(1770)$.
A small peak has been observed at 1750 MeV for $J=0$ in two sets of data on
$\pi \pi \to K_SK_S$ [11,12].
Secondly, a partial wave analysis of $J/\psi \to \gamma (4\pi)$ has fitted $4\pi$
peaks at 1780 MeV with $J=0$ \cite {13}.
The GAMS collaboration has reported \cite {14} an $X(1740)$ decaying to $\eta \eta$
at $1744 \pm 15$ MeV, but they find a width $<90$ MeV.
Their angular distribution is flat, but spin 0 is not claimed because of 
uncertainty about the production mechanism.
E760 data \cite {15} on $\bar pp \to \eta \eta \pi ^0$ show peaks at 1500, 
$1748 \pm 10$ and 2100 MeV; 
they do not report a spin-parity analysis.
In view of the fact that we observe strong production of $f_0(1500)$ and $f_0(2100)$
in our $\eta \eta \pi ^0$ data, it seems plausible that their peak at $\sim 1750$
MeV may be the same resonance as we report here.
Their width of $264 \pm 25$ MeV is acceptably close to ours.
The BES group observes an $f_0$ at 1781 MeV in $J/\psi \to \gamma (K^+K^-)$
\cite {16}.

Two $0^+$ $\bar qq$ resonances are to be expected in this mass range.
One is the radial excitation of $f_0(1300-1380)$.
The second is the missing $s\bar s$ state from a $0^+$ nonet made up of
$f_0(1300-1380)$, $a_0(1450)$ and $K^*(1430)$.
The resonance we observe could be either of these or a linear combination.
The L3 collaboration \cite {17} has reported a narrow signal at $1793\pm 18$ MeV
in $K_SK_S$, but without a spin-parity determination.
A more recent analysis \cite {18} gives $M=1770 \pm 20$ MeV, $\Gamma = 235 \pm 47$
MeV, very close to values reported here.

There is a long-standing debate over the spin of $f_J(1710)$, reported by the
Particle Data Group as having a mass of $1712 \pm 8$ MeV and a width of
$133 \pm 14$ MeV.
The mass we observe, 1770 MeV, seems too far removed from 1712 MeV to be the same
resonance.
It seems likely that there is one resonance at 1710 MeV with $J=0$ or 2 and a
second at 1770 MeV.
This is what the BES collaboration found \cite{16}.
There is then the possibility that the $f_J(1710)$ is also weakly present in our
data. 
At 900 MeV/c, there is a small surplus of events in the $\eta \eta$ projection of
Fig. 1(f) from 1600 to 1700 MeV.
At 1200 MeV/c there is again some indication of extra events in four bins close to
1700 MeV.
Our fits include the 500 MeV wide $f_2(1980)$ demanded by the data of Ref. [1]
at higher beam momenta.
In ref. [1], the $2^+$ signal is visible over the whole mass range from 1550 to
2000 MeV.
The $\cos \alpha$ dependence of Fig. 3(a) at 1200 MeV/c is again clear
evidence for the presence of non-zero spin, but it does not appear to be
narrowly concentrated in mass.
Again, it is to be associated with $f_2(1980)$.
At lower beam momenta, where the phase space is limited, there is the possibility
of confusion between $f_2(1980)$ and $f_2(1710)$. 
At all beam momenta, we have tried adding to the present fit $f_0(1712)$ or
$f_2(1712)$ with the PDG width of 133 MeV.
Improvements to the fit are listed in columns (c) and (d) of Table 3
respectively.
They are not significant, compared with our test with fake signals; 
they make no visible improvement to mass projections.
If one tries to fit the extra events of Fig. 1(h) with $f_J(1710)$, the optimum
mass is 1690 MeV but the width required is $\le 40$ MeV and optimises at 25 MeV.
This width is so much less than the PDG value for $f_J(1710)$ that we have no
confidence in the presence of an additional $f_J(1710)$.
An upper limit ($90\%$ confidence level) on its branching ratio at all momenta
is $4\%$.

In summary, there is strong $(>8\sigma)$ evidence at each of three momenta for
an $f_0$ with $M = 1770 \pm 12$ MeV and $\Gamma = 220 \pm 40$ MeV.
There is no significant evidence for a further resonance at 1710 MeV.

\vskip 4mm
{\bf Acknowledgements}
\vskip 2mm
We thank the Crystal Barrel Collaboration for allowing use of the data.
We wish to thank the technical staff of the LEAR machine group
and of all the participating institutions for their invaluable
contributions to the success of the experiment.  
We acknowledge financial support from the British Particle Physics and
Astronomy Research Council (PPARC) and the U.S. Department of Energy.
The St. Petersburg group wishes to acknowledge financial support from PPARC and
INTAS grant RFBR 95-0267.


\begin {thebibliography}{99}
\bibitem {1} A.V. Anisovich  et al, to be submitted to Physics Letters.
\bibitem {2} Particle Data Group, Euro. Phys. J. C3 (1998) 1.
\bibitem {3} V.V. Anisovich et al., Phys. Lett. B 323 (1994) 233.
\bibitem {4} C. Amsler et al., Phys. Lett. B 355 (1995) 425.
\bibitem {5} A. Abele et al., Nucl. Phys. A 609 (1996) 562.
\bibitem {6} C. Amsler et al., Phys. Lett. B 322 (1994) 431.
\bibitem {7} C. Amsler et al., Phys. Lett. B 333 (1994) 277.
\bibitem {8} D. Barberis et al., Phys. Lett. B 413 (1997) 225.
\bibitem {9} D.V. Bugg, V.V. Anisovich, A. Sarantsev, B.S. Zou,
Phys. Rev. D 50 (1994) 4412.
\bibitem {10} A.V. Anisovich et al., to be submitted to  Phys. Lett. B.
\bibitem {11} B.V. Bolonkin et al., Nucl. Phys. B 309 (1988) 426.
\bibitem {12} A. Etkin et al., Phys. Rev. D 25 (1982) 1786, 2446.
\bibitem {13} D.V. Bugg et al., Phys. Lett. B 353 (1995) 378.
\bibitem {14} D. Alde et al., Phys. Rev. B 284 (1992) 457.
\bibitem {15} T.A. Armstrong et al., Phys. Lett. B 307 (1993) 394.
\bibitem {16} J.Z. Bai et al., Phys. Rev. Lett. 77 (1996) 3959.
\bibitem {17} M. Acciarri et al., Phys. Lett. B 363 (1995) 118.
\bibitem {18} S. Braccini, XXIX Int. Conf. on High Energy Physics, 
Vancouver, 1998.  
\end {thebibliography}
\end{document}